\documentstyle[twocolumn,prl,epsf,aps]{revtex}

\begin{document}

\draft

\title{Spin and charge excitations
in incommensurate spin density waves}

\author{Eiji Kaneshita, Masanori Ichioka and Kazushige Machida}
\address{Department of Physics, Okayama University,
         Okayama 700-8530, Japan}
\date{\today}

\maketitle

\begin{abstract}
Collective excitations both for spin- and charge-channels 
are investigated in incommensurate spin density wave (or stripe) states on two-dimensional Hubbard model. 
By random phase approximation, the dynamical susceptibility 
$\chi({\bf q},\omega)$ is calculated for full range of $({\bf q},\omega)$ 
with including all higher harmonics components. 
An intricate landscape of the spectra in $\chi({\bf q},\omega)$ is obtained.
We discuss the anisotropy of the dispersion cones for spin wave excitations, 
and for the phason excitation related to the motion of the stripe line. 
Inelastic neutron experiments 
on Cr and its alloys and stripe states of 
underdoped cuprates are proposed. 
\end{abstract}


\narrowtext


Recently a remarkable series of elastic neutron experiments has been 
performed on La$_{2-x}$Sr$_{x}$CuO$_{4}$~\cite{LSCO}.
It reveals static incommensurate spin density wave (ISDW) structure 
in underdoped regions.  
The superconducting samples show that
the modulation vector $\bf Q$ are characterized by 
${\bf Q}=2\pi(\frac{1}{2},\frac{1}{2}\pm\delta)$ or  
$2\pi(\frac{1}{2}\pm\delta,\frac{1}{2})$ in momentum space. 
The incommensurate modulation runs vertically in the $a$- or $b$-axis 
of the CuO$_2$ plane. 
The static stripe structures have been observed also in superconducting 
(La, Nd)$_{2-x}$Sr$_{x}$CuO$_{4}$~\cite{LSNCO} and 
insulating La$_{2-x}$Sr$_{x}$NiO$_{4}$~\cite{LSNiO}. 
Since the establishment of static orders on
these systems, 
the study of spin and charge dynamics; $\chi({\bf q},\omega)$ 
has been just started experimentally. 

Prior to these studies on cuprates, ISDW is well known in itinerant electron systems, 
such as a typical example Cr and its alloys for long time.  
While its static properties are fairly well 
understood, 
its dynamical properties remain largely unexplored~\cite{MachidaFujita,Fawcett}. 
A part of reasons stems from the fact that the theoretical description 
is underdeveloped, thus the interplay between theory and experiment 
was not satisfactory\cite{Hayden}. 
The intensive efforts for extracting spin and charge 
excitations in Cr and its alloys and also high $T_{\rm c}$ cuprates are strongly motivated by 
a hope that knowing the collective excitations or fluctuation spectrum in strongly correlated systems
may lead us to a clue to understanding the mechanism of high $T_{\rm c}$
superconductivity. 

There exist a lot of theoretical works on dynamical 
properties, beginning from
a seminal paper by Fedders and Martin\cite{Fedders}
to recent work by Fishman and Liu\cite {Fishman}.
The former  first derives spin wave mode in the transverse spin 
excitation of itinerant electron systems. 
The latter investigates various transverse 
and longitudinal modes at the ${\bf Q}$ position,
based on a one-dimensional (1D) model within 
the random phase approximation (RPA). 
Their theory takes account of only fundamental order parameter 
$\Delta_{\bf Q}$, neglecting higher harmonics
$\Delta_{2{\bf Q}}$, $\Delta_{3{\bf Q}}$, $\cdots$ associated with the
incommensurability $\delta$. 
By extending their work~\cite{Ichioka1D}, 
we calculate the dynamical spin and charge susceptibilities 
$\chi({\bf q},\omega)$ in the full range of 
the two dimensional (2D) wave number ${\bf q}$
for entire Brillouin zone
and the excitation energy $\omega$ up to the band width. 
We take account of all the possible higher harmonics. 
This will turn out to be extremely crucial in correctly evaluating these quantities. 
The multi-dimensional calculation here allows 
us to extract a wealth of information on stripe motions such as translation, or meandering, etc. and on the
anisotropy of excitation cones. This kind of calculation
has not been done before to our knowledge.

We start with the Hubbard model on 2D square lattice:
$H=-t\sum_{i,j,\sigma}C^{\dagger}_{i,\sigma} 
C_{j,\sigma}+U\sum_in_{i\uparrow}n_{i\downarrow}$. 
To calculate $\chi({\bf q},\omega)$, 
we first set up the incommensurate SDW ground state. 
Assuming a periodic spin and associated charge orderings, 
we introduce the order parameter 
$\langle n_{i,\sigma} \rangle =\sum_l e^{{\rm i}l{\bf Q}\cdot{\bf r_i}}
\langle n_{l{\bf Q},\sigma} \rangle$ with $\delta=1/N$. 
In $N$-site periodic case, the Brillouin zone is reduced to $1/N$-area. 
and energy dispersion is split to $N$ bands. 
We write ${\bf k}={\bf k}_0 + m{\bf Q}$ ($m=0,\ 1,\ \cdots,\ N-1$), where 
${\bf k}_0$ is restricted within the  reduced Brillouin zone. 
Then the Hamiltonian is reduced to 
\begin{eqnarray} &&
H=\sum_{{\bf k}_0,\sigma}
\sum_{m,n}C^{\dagger}_{{\bf k}_0+m{\bf Q},\sigma}
({\hat H}_{{\bf k}_0,\sigma})_{m,m'}C_{{\bf k}_0+m'{\bf Q},\sigma} 
\nonumber \\ && 
=\sum_{{\bf k}_0,\sigma,\alpha}E_{{\bf k}_0,\sigma,\alpha}
\gamma^\dagger_{{\bf k}_0,\sigma,\alpha} 
\gamma_{{\bf k}_0,\sigma,\alpha}. 
\label{Hamiltonian1}
\end{eqnarray} 
The $N \times N$ Hamiltonian matrix 
$ ( {\hat H}_{{\bf k}_0,\sigma}  )_{m,m'}
=\epsilon({\bf k}_0+m{\bf Q})\delta_{m,m'}
+U \langle n_{(m-m'){\bf Q},-\sigma} \rangle $
is diagonalized by a unitary transformation 
$C_{{\bf k}_0+m{\bf Q},\sigma}=\sum_{\alpha} 
u_{{\bf k}_0,\sigma,\alpha,m}\gamma_{{\bf k},\sigma,\alpha}$. 
The calculation is iterated until all
the order parameters satisfy 
the self-consistent condition 
$\langle n_{lQ\sigma} \rangle = N_k^{-1}
\sum_{k_0, m, \alpha} u^{\ast}_{{\bf k}_0,\sigma,\alpha, m}
u_{{\bf k}_0,\sigma,\alpha, m+l}f(E_{{\bf k}_0,\sigma, \alpha})  $. 
Here, $N_k=\sum_{{\bf k}_0,m}1$. 

We construct the thermal Green function as 
\begin{equation}
g_\sigma({\bf r},{\bf r}', {\rm i} \omega_n)=\sum_{{\bf k}_0,\alpha}
\frac{u_{{\bf k}_0,\sigma,\alpha}({\bf r})
u^\ast_{{\bf k}_0,\sigma,\alpha}({\bf r'})}
{ {\rm i} \omega_n - E_{{\bf k}_0,\sigma,\alpha}},
\label{eq:Gf} 
\end{equation}
with
$u_{{\bf k}_0,\sigma,\alpha}({\bf r}_i)=N_k^{-1/2} \sum_m
{\rm e}^{{\rm i}({\bf k}_0+m{\bf Q})\cdot {\bf r}_i}
u_{{\bf k}_0,\sigma,\alpha,m}$, 
and evaluate the dynamical susceptibilities; 
the spin longitudinal mode $\chi_{zz}({\bf q},\omega)
=\langle\langle S_z;S_z \rangle\rangle_{{\bf q},\omega}$, 
the transverse one $\chi_{xx}({\bf q},\omega)= 
\langle\langle S_x;S_x \rangle\rangle_{{\bf q},\omega}$, 
and the charge susceptibility $\chi_{nn}({\bf q},\omega)=
\langle\langle n;n \rangle\rangle_{{\bf q},\omega}$. 
The Fourier transformation of Eq. (\ref{eq:Gf}) is given by 
\begin{eqnarray} &&
g_\sigma({\bf k}+l_1 {\bf Q},{\bf k}+l_2 {\bf Q}, {\rm i} \omega_n)
\nonumber \\ && 
=\frac{1}{N_k}\sum_{{\bf r}_1,{\bf r}_2}
e^{-{\rm i}({\bf k}+l_1{\bf Q})\cdot{\bf r}_1}
e^{ {\rm i}({\bf k}+l_2{\bf Q})\cdot{\bf r}_2}
g_\sigma({\bf r}_1,{\bf r}_2, {\rm i} \omega_n)  
\nonumber \\ && 
=\sum_\alpha
\frac{u_{{\bf k},\sigma,\alpha,l_1}
 u^\ast_{{\bf k},\sigma,\alpha,l_2} }
{ {\rm i} \omega_n - E_{{\bf k},\sigma,\alpha}},
\label{eq:4.3}
\end{eqnarray}
where $l_1, \ l_2=0, \ 1, \cdots, N-1$.
In the presence of the order parameter $\langle n_{l{\bf Q},\sigma} \rangle$,
the incoming- and outgoing-momentum of $g_\sigma$ can differ
by $l{\bf Q}=(l_1 -l_2){\bf Q}$. 
Then $g_\sigma({\bf k}+l_1 {\bf Q},{\bf k}+l_2 {\bf Q}, {\rm i} \omega_n)$ 
is an $N \times N$ matrix with indexes $l_1$ and $l_2$. 
The bare susceptibility is given by 
$\chi_0^{\sigma \sigma'} ({\bf r}_1 ,{\bf r}_2, {\rm i} \Omega_n)
=-T  \sum_{\omega_n}
g_\sigma ({\bf r}_1,{\bf r}_2, {\rm i} \omega_n + {\rm i} \Omega_n)
g_{\sigma'} ({\bf r}_2,{\bf r}_1, {\rm i} \omega_n ). $
Its Fourier transformation 
$\chi_0^{\sigma \sigma'} 
({\bf k}+l_1 {\bf Q},{\bf k}+l_2 {\bf Q}, {\rm i} \omega_n) $ 
is also $N \times N$ matrix~\cite{Lee}. 
We use the analytic continuation 
${\rm i}\omega_n \rightarrow \omega + {\rm i}\eta$. 
Typically, we use $\eta=0.001t$ in our numerical calculation. 

The RPA equation for $\chi_{S_\uparrow S_\downarrow}
= \langle\langle S_\uparrow; S_\downarrow \rangle\rangle$ is written as 
\begin{eqnarray} &&
\chi_{S_\uparrow S_\downarrow}({\bf r}_1,{\bf r}_3, \omega)
=\chi_0^{\uparrow \downarrow} ({\bf r}_1,{\bf r}_3, \omega)
\nonumber \\ &&
+ U \sum_{{\bf r}_2} \chi_0^{\uparrow \downarrow} 
({\bf r}_1,{\bf r}_2, \omega)
\chi_{S_\uparrow S_\downarrow}({\bf r}_2 ,{\bf r}_3 , \omega) .
\label{eq:Dyssr}
\end{eqnarray}
After Fourier transformation to ${\bf k}$-space, 
Eq. (\ref{eq:Dyssr}) is reduced to a matrix equation of 
$\chi_0^{\uparrow \downarrow}$ and $\chi_{S_\uparrow S_\downarrow}$. 
By solving it, we obtain 
$\chi_{xx}({\bf k}+l_1 {\bf Q},{\bf k}+l_2 {\bf Q}, \omega)$. 
In a similar manner, we calculate 
$\langle\langle n_\uparrow;n_\uparrow \rangle\rangle$ and 
$\langle\langle n_\uparrow;n_\downarrow \rangle\rangle$, and obtain 
$\chi_{zz}({\bf k}+l_1 {\bf Q},{\bf k}+l_2 {\bf Q}, \omega)$ 
and $\chi_{nn}({\bf k}+l_1 {\bf Q},{\bf k}+l_2 {\bf Q}, \omega)$. 
The neutron scattering experiments observe the imaginary part of 
the dynamical susceptibility 
$\chi''({\bf q},\omega)={\rm Im}\chi({\bf q},{\bf q},\omega)$. 
Since the signal is observed as spatial average, 
we consider the diagonal part of 
$\chi({\bf k}+l_1 {\bf Q},{\bf k}+l_2 {\bf Q}, \omega)$.

According to standard linear response theory, 
the spatio-temporal oscillation pattern of a collective mode can be analyzed 
by $\chi$.
In the presence of an infinitesimal external field
$h_{O'}({\bf k'},\omega)$ coupled to an operator $O'(=S_x,\ S_z,\ n)$, 
the response of the operator $O$ is given by
$\delta \langle O({\bf k}_0 +l_1 {\bf Q},\omega) \rangle
=\sum_{l_2} \chi_{OO'}({\bf k}_0 +l_1 {\bf Q},{\bf k}_0 +l_2 {\bf Q},\omega)
h_{O'}({\bf k}_0 +l_2 {\bf Q},\omega) .$
When the external field is a plane wave;
$h_{O'}({\bf r},t)
=\bar{h}_{O'}{\rm e}^{{\rm i}({\bf q}\cdot{\bf r}- \omega t)}$
with a small amplitude $\bar{h}_{O'}$,
the response is given by
\begin{eqnarray} &&
\delta \langle O({\bf r},t) \rangle
= \sum_l \chi_{OO'}({\bf q}+l{\bf Q},{\bf q},\omega)
{\rm e}^{{\rm i}({\bf q}+l{\bf Q})\cdot{\bf r} }
\bar{h}_{O'}{\rm e}^{- {\bf i}\omega t}.
\nonumber \\
\label{eq:LRp}
\end{eqnarray}

\begin{figure}
\begin{center}
\leavevmode
\epsfxsize=40mm
\epsfbox{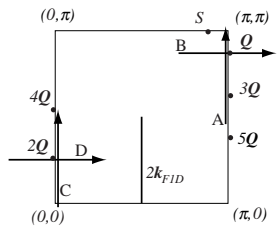}
\end{center}
\caption{
The paths A-D along which we show $\chi({\bf q},\omega)$ 
in the momentum space. 
The point ${\bf Q}$ is the ordering vector. 
$2{\bf Q},\ 3{\bf Q},\cdots$ are its higher harmonics points.  
${\bf S}$ is the silent position. 
The line $2{\bf k}_{\rm F1D}$ shows the nesting wave number of the 
1D Fermi surface in the metallic stripe state.
}
\label{fig:path}
\end{figure}

We consider the vertical stripe case for $U=4t$ and the hole density $n_{\rm h}=1/8$ 
as a representative case
(The enrgy is scaled by $t$ from now on). 
As we do not include the nearest neighbor hopping $t'$, 
the lowest energy ground state is an SDW-gapped insulator 
with $\delta=n_{\rm h}/2$, i.e. $N=16$. 
The detailed ground state properties are 
reported previously~\cite{Machida,Ichioka}.  
For our parameters, the single particle SDW gap $E_g=0.41$.  
In the ISDW state, 
the spatial profile is characterized by a distorted sinusoidal, or soliton form with a midgap band. 
The higher harmonics are determined as 
$|M_{l{\bf Q}}/M_{\bf Q}|=$0.08  ($l=3$), 0.005 ($l=5$), 
$2.3 \times 10^{-3}$ ($l=7$), 
where $M_{l{\bf Q}}= \langle n_{l{\bf Q},\uparrow} \rangle 
- \langle n_{l{\bf Q},\downarrow} \rangle$. 
In the limit of the half-filling $(N \rightarrow \infty)$, 
the ratio $|M_{(2n+1){\bf Q}}/M_{\bf Q}|$ increases and approaches 
$(2n+1)^{-1}$, since the profile of the spin structure 
approaches square wave form~\cite{Machida,Ichioka}.
The structure factor has spots at $(2n+1){\bf Q}$ in the 
spin structure, and at $2n{\bf Q}$ in the charge structure. 
These spots are observed by the elastic neutron scattering in Cr and its alloys~\cite{Fawcett}. 
Their positions in the momentum space are shown in Fig. \ref{fig:path}.

\begin{figure}
\begin{center}
\leavevmode
\mbox{(a)
\epsfxsize=35mm
\epsfbox{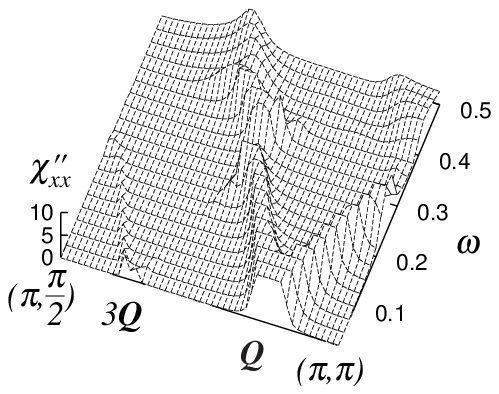}
(b)
\epsfxsize=35mm
\epsfbox{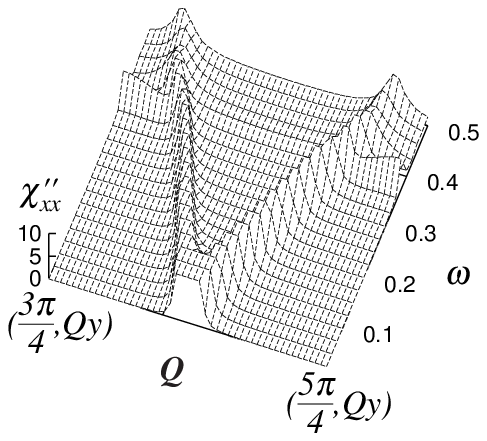}}
\end{center}
\caption{
Spin transverse mode $\chi''_{xx}({\bf q},\omega)$ along path A (a) and B (b). 
$\omega$ is scaled by $t$. 
We cut off the peak height for $\chi''_{xx}>10$ to show the 
low intensity structure. 
}
\label{fig:chixx}
\end{figure}

Let us start with the excitation of the spin transverse mode. 
In Fig. \ref{fig:chixx}, we show $\chi''_{xx}({\bf q},\omega)$ along 
paths A and B of Fig. \ref{fig:path}.  
The gapless spin wave modes emanate not only ${\bf Q}$, 
but also from $3{\bf Q}$ and other odd harmonics.  
The ridge of $\chi''({\bf q},\omega)$ shows singularity reflecting 
the dispersion of the collective mode. 
These modes at $(2n+1){\bf Q}$ have an identical dispersion relation, 
because every $(2n+1){\bf Q}$-modes couple each other in the RPA 
equation (\ref{eq:Dyssr}).  
The same dispersion pattern appears in each reduced Brillouin zone. 
But their intensities are different. 
With increasing $l$, the intensity is decreased as follows,  
$\chi''_{xx}(l{\bf Q},0.025t)/
  \chi''_{xx}({\bf Q},0.025t)= 6.5 \times 10^{-3}$  ($l=3$), 
$2.7 \times 10^{-5}$ ($l=5$), $5.8 \times 10^{-6}$ ($l=7$). 
It is found that the intensity ratio is obeyed 
\begin{equation} 
\frac{\chi''_{xx}((2n+1){\bf Q},\omega\rightarrow0)}
     {\chi''_{xx}({\bf Q},\omega\rightarrow0)}
\sim \left[ \frac{M_{(2n+1){\bf Q}}}{M_{{\bf Q}}} \right]^2 . 
\label{chi-ratio}
\end{equation}
With approaching the half-filling ($N\rightarrow\infty$)
the above ratio is expected to increase and approach $(2n+1)^{-2}$. 
Then, the higher harmonics spot at $(2n+1){\bf Q}$ may have enough intensity to 
be observed near half-fillings such as in Cr or underdoped cuprates. 

We analyze the oscillation pattern by Eq. (\ref{eq:LRp}). 
The response of $\delta \langle S_x({\bf r},t) \rangle $ shows 
the same spin pattern as $S_z$ of the ground state at 
$\omega \sim 0$ for  $(2n+1){\bf Q}$. 
It means that the ground state spin structure is rotated as it is without modulation, 
since it is a Goldstone mode. 
The external field of the wave number $(2n+1){\bf Q}$ 
couples to $M_{(2n+1){\bf Q}}$
and makes the ground state spin structure rotate. 
Then, we can conclude that the spin transverse mode is a spin wave. 
With increasing $\omega$ along the dispersion curve, 
the spin wave oscillation shows the deviation from the spin pattern 
of the ground state, reflecting the wave number of the external field. 

The reconnection of the dispersion curve occurs at the 
reduced Brillouin zone boundary. 
Then, instead of the simple intersection of two dispersions, the small gap appears at 
$\omega \sim 0.3t$ in Fig. \ref{fig:chixx}. 
With increasing $\omega$, the intensity of $\chi''_{xx}({\bf q},\omega)$ 
decreases as $1/\omega$ along the dispersion curve, 
except for the weakened intensity at the gap position. 
For $\omega > E_g$, there exists other modes reflecting 
the fluctuation of the magnetic moment amplitude. 
It is a character of itinerant magnets and absent 
in localized spin magnets.

\begin{figure}
\begin{center}
\leavevmode
\epsfxsize=45mm
\epsfbox{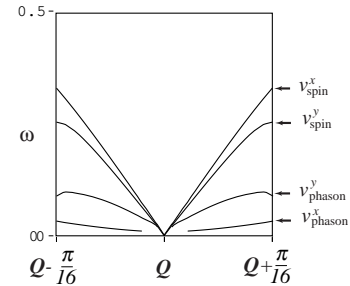}
\end{center}
\caption{
Dispersion curve of the spin-wave excitation 
($v^x_{\rm spin}$ for path B and $v^y_{\rm spin}$ for path A) and 
phason excitation 
($v^x_{\rm phason}$ for path B and $v^y_{\rm phason}$ for path A). 
The slope at $\omega\sim0$ gives velocity of each mode. 
}
\label{fig:velocity}
\end{figure}

Figure \ref{fig:velocity} shows the dispersion curve along the 
path A ($q_y$-direction)  and B ($q_x$-direction) near ${\bf Q}$ in Figs.  \ref{fig:chixx} (a) and (b). 
The spin wave velocity $v^x_{\rm spin}$ ($v^y_{\rm spin}$) is defined 
by the slope of the $q_x$- ($q_y$-) direction at $\omega \sim 0$. 
The spin modulation parallel to the stripe (domain wall) corresponds 
to $v^x_{\rm spin}$. 
In this direction, the staggered spin moment has a constant amplitude. 
The modulation perpendicular to the stripe corresponds 
to $v^y_{\rm spin}$. 
In this direction,  the spin moment is modulated and
suppressed when it crosses the 
stripe region. 
In Fig. \ref{fig:velocity}, $v^x_{\rm spin} > v^y_{\rm spin}$. 
It indicates that the spin modulation is easier for 
the direction perpendicular to the stripe. 
In other words, the effective exchange integral $J$ 
across the stripe becomes weaker than that parallel to the stripe. 
It is the first time to microscopically derive the anisotropy of the 
spin wave velocity.
As $U$ increases, $v_{\rm spin}$ decreases. 
These results are reasonable in view of the correspondence between Hubbard and Heisenberg models: 
$v_{\rm spin} \sim J\propto t^2 / U$. 
We have done the same calculation for the diagonal stripe to confirm that the spin wave velocity is similar.

\begin{figure}
\begin{center}
\leavevmode
\mbox{(a)
\epsfxsize=35mm
\epsfbox{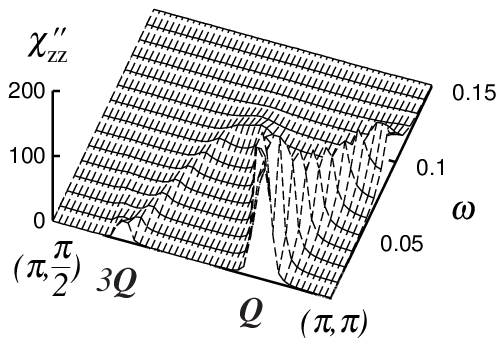}
(b)
\epsfxsize=35mm
\epsfbox{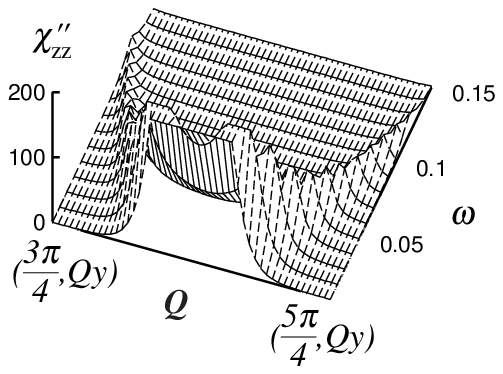}}
\end{center}
\caption{
Spin longitudinal mode $\chi''_{zz}({\bf q},\omega)$ 
along path A (a) and B (b).
We cut off the peak height for $\chi''_{zz}>200$. 
}
\label{fig:chizz}
\end{figure}

The longitudinal spin mode $\chi''_{zz}({\bf q},\omega)$ is shown 
in Fig. \ref{fig:chizz} along paths A and B of Fig. \ref{fig:path}. 
Low energy modes appears at $(2n+1){\bf Q}$. 
Along path A, the dispersion relation is continuous 
and repeated, touching at $(2n+1){\bf Q}$. 
Away from ${\bf Q}$, the intensity decreases as 
$\chi''_{zz}(l{\bf Q},0.01t)/
  \chi''_{zz}({\bf Q},0.01t)= 5.9 \times 10^{-2}$  ($l=3$), 
$6.9 \times 10^{-4}$ ($l=5$), $5.3 \times 10^{-4}$ ($l=7$). 
These ratios are larger than those given by Eq. (\ref{chi-ratio}), which is, thus, not satisfied for 
$\chi_{zz}$.  
The intensity $\chi''_{zz}({\bf q},\omega)\sim \frac{1}{3}
\chi''_{xx}({\bf q},\omega)$ near ${\bf Q}$ along each dispersion 
relation in our results. 
There, $\chi''_{zz}({\bf q},\omega)\sim 1/\omega$ with increasing $\omega$. 

\begin{figure}
\begin{center}
\leavevmode
\mbox{(a)
\epsfxsize=35mm
\epsfbox{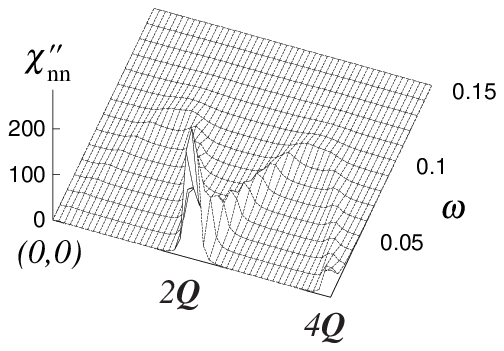}
(b)
\epsfxsize=35mm
\epsfbox{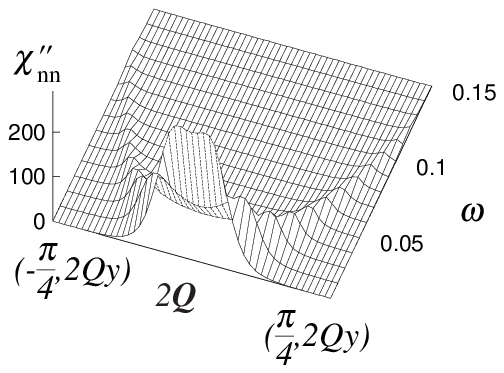}}
\end{center}
\caption{
Charge mode $\chi''_{nn}({\bf q},\omega)$ 
along path C (a) and D (b).
}
\label{fig:chinn}
\end{figure}

The charge mode ${\chi''_{nn}({\bf q},\omega)}$ is shown in 
Fig. \ref{fig:chinn} along paths C and D of Fig. \ref{fig:path}. 
The low energy mode appears at $2n{\bf Q}$. 
It has the identical dispersion curve as that of $\chi_{zz}$, 
because $S_z$ and $n$ couple each other in the RPA equation. 
As in Fig. \ref{fig:chizz} (a), it is a continuous curve along path C 
and its intensity deceases away from $2{\bf Q}$. 
The $2{\bf Q}$ mode comes from even harmonics, 
as $\langle n_{2{\bf Q},\sigma} \rangle$ describes the charge modulation in 
the ISDW. 

We analyze the oscillation pattern of $S_z$ and $n$ by Eq. (\ref{eq:LRp}). 
Along the dispersion curve of $\chi_{zz}$ of Fig. \ref{fig:chizz} 
($\chi_{nn}$ of Fig. \ref{fig:chinn}), the response is enhanced by the 
resonance with the external field coupled to $S_z$ ($n$). 
The response of $\delta \langle S_z({\bf r},t) \rangle $ and 
$\delta \langle n({\bf r},t) \rangle $ shows large amplitude 
near the stripe region. 
It means that this excitation is related to the motion of the stripe, 
i.e., phason mode. 
This analysis shows that the collective mode at $\omega \sim 0$ 
is the translational motion, where whole stripes move together to 
the same direction. 
When the pinning (such as the energy difference between the site-centered 
stripe and the bond-centered stripe) is negligible, 
this translational mode is a Goldstone mode and gapless as in 
Figs \ref{fig:chizz} and \ref{fig:chinn}. 
From the analysis of Eq. (\ref{eq:LRp}), we understand that the 
excitation along the $q_y$-direction is a compress mode 
and that along the $q_x$-direction is a meandering mode. 
In the compress mode, 
the inter-stripe distance is modulated periodically in the direction 
perpendicular  to the stripe, with keeping the straight line shape. 
In the meandering mode, each stripe meanders along the stripe direction 
with keeping the same inter-stripe distance. 
From Fig. \ref{fig:velocity}, we see $v^x_{\rm phason}<v^y_{\rm phason}$. 
It means the meandering motion is easier to occur compared with the 
compress mode in the vertical stripe case of this model. 
As $N$ increases, $v^x_{\rm phason}$ and $v^y_{\rm phason}$ decrease. 
It is because effective interaction between neighbor stripes is weak 
and each stripe can move more freely when the inter-stripe distance becomes long. 
This slow velocity of the longitudinal mode for large $N$ 
may be related to the un-identified Fincher-Burke mode~\cite{Fincher} observed in Cr. 
This identification deserves further experimental and 
theoretical studies.

We are also interested in the silent position 
${\bf q}={\bf S}=2\pi(\frac{1}{2}\pm \delta, \frac{1}{2})$, 
which is an equivalent position to ${\bf Q}$ in the paramagnetic state 
above  N\'{e}el temperature (see Fig.\ref{fig:path}). 
This silent mode is related to a critical scattering in Cr~\cite{Boni}. 
There exists a large intensity in $\chi''_{xx}({\bf S},\omega)$, 
whose dispersion has a gap of the order $E_g$. 
In $\chi''_{zz}({\bf q},\omega)$, the excitation at ${\bf S}$ 
(but slightly shifted to lower $q_x$) is shifted to lower energy 
as $N =8 \rightarrow 12 \rightarrow 16$. 
It almost touches $\omega =0$ for $N=16$. 
It may suggest that we are approaching the transition to another 
low energy ground state (such as diagonal stripe) 
for lower $n_h$~\cite{Machida,Ichioka}. 

So far we mention only the insulating stripe state.
The corresponding metallic stripe state can be also stabilized 
by merely introducing the next nearest hopping $t'$. 
The lowest energy state is given by $\delta\sim n_{\rm h}$, 
and the Fermi level situates in the so-called midgap band\cite{Ichioka}. 
Then, we set $N=8$ in the metallic case. 
Since $v^y_{\rm spin}$ increases and 
$v^x_{\rm spin}$ decreases, we find $v^y_{\rm spin} > v^x_{\rm spin}$ in $\chi''_{xx}$.  
There is low energy excitation also at $\bf S$. 
In $\chi''_{zz}$ and $\chi''_{nn}$, 
the excitation at ${\bf Q}$ or $2{\bf Q}$ has a gap. 
The low energy excitation appears along the line at $q_x = \pi/2$ 
which is presented by a line in Fig. \ref{fig:path}. 
It is the 1D CDW or SDW fluctuation mode within the stripe line, 
and originated from the Fermi surface nesting 
$2{\bf k}_{\rm F1D}$ of the parallel 1D Fermi lines 
(Fig. 12(d) in Ref. \onlinecite{Ichioka}) of the stripe 
state~\cite{Zhou}. 
Since the 1D Fermi state has a gap near $(\frac{\pi}{2},\frac{\pi}{2})$, 
the intensity of $\chi''(2{\bf k}_{\rm F1D},\omega\sim0)$ vanishes near 
$(\frac{\pi}{2},\pi)$. 
These low energy excitations are diffusive since $E_g=0$ in 
metallic state.

In summary, we have investigated the dynamical susceptibilities
of transverse and longitudinal spin channels and
charge one for the whole space spanned by 2D wave vector $\bf q$
and the energy $\omega$, and identified several elementary
excitations; the spin wave mode and the 
phason mode related to the motion of the stripe line.
This allows us to construct the whole landscape of 
$(\bf q,\omega)$ space for the excitation spectra of various channels:
The identical dispersion relation is replicated at every $(2n+1){\bf Q}$, 
which has the anisotropic excitation cones along $q_x$- and 
$q_y$-directions. 
Our predictions about $\chi({\bf q},\omega)$ are directly testable 
by careful inelastic neutron experiments on Cr alloys and underdoped cuprates.

We thank G. Shirane, Y. Endoh, K. Yamada, T. Fukuda, M. Wakimoto, 
M. Matsuda and M. Fujita for their useful discussions 
and  information.



\begin{references}


\bibitem{LSCO}
M. Matsuda, {\it et al.}, cond-mat/0003466. 
S. Wakimoto, {\it et al.}, Phys. Rev. B {\bf 60}, 769 (1999).
T. Suzuki, {\it et al.}, Phys. Rev. B {\bf 57}, R3229 (1998). 


\bibitem{LSNCO}
J.M. Tranquada, {\it et al.}, Phys. Rev. Lett. {\bf 78}, 338 (1997). 

\bibitem{LSNiO}
J.M. Tranquada, {\it et al.}, Phys. Rev. B {\bf 54},  12318 (1996). 
H. Yoshizawa, {\it et al.}, Physica B {\bf 241-243}, 880 (1998).



\bibitem{MachidaFujita}
K. Machida and M. Fujita, Phys. Rev. B{\bf 30}, 5284 (1984).

\bibitem{Fawcett}
For reviews, 
E. Fawcett, Phys. Mod. Phys. {\bf 60}, 209 (1988).
E. Fawcett, {\it et al.}, Phys. Mod. Phys. {\bf 66}, 26 (1994).

\bibitem{Hayden}
For more recent neutron experiments on Cr alloys, see
S.M. Hayden, {\it et al.}, Phys. Rev. Lett. {\bf 84}, 999 (2000) 
and references therein.

\bibitem{Fedders}
P. A. Fedders and P.C. Martin, Phys. Rev. {\bf 143}, 1845 (1966).

\bibitem{Fishman}
R.S. Fishman and S.H. Liu, Phys. Rev. Lett. {\bf 76}, 2398 (1996); 
Phys. Rev. B {\bf 54}, 7233 and 7252 (1996). 

\bibitem{Ichioka1D} 
M. Ichioka, E. Kaneshita, and K. Machida, in preparation. 
The details of our formulation and analysis are described in the 1D case.


\bibitem{Lee}
P.A. Lee, T.M. Rice and P.W. Anderson, Solid State Commun. 
{\bf 14}, 703 (1974).

\bibitem{Machida}
K. Machida, Physica C{\bf 158}, 192 (1989).
M. Kato, {\it et al.}, J. Phys. Soc. Jpn., {\bf 59}, 1047 (1990).
K. Machida and M. Ichioka,  J. Phys. Soc. Jpn., {\bf 68}, 2168 (1999).

\bibitem{Ichioka}
M. Ichioka and K. Machida,  J. Phys. Soc. Jpn., {\bf 68}, 4020 (1999).

\bibitem{Fincher}
C.R. Fincher, Jr., {\it et al.}, Phys. Rev. B{\bf 24}, 1312 (1981).
S.K. Burke, {\it et al.}, Phys. Rev. Lett. {\bf 51}, 494 (1983).

\bibitem{Boni}
P. B\"oni, {\it et al.}, Phys. Rev. Lett. {\bf 51}, 494 (1983).

\bibitem{Zhou}
X.J. Zhou, {\it et al.}, Science {\bf 286}, 268 (1999). 
A. Ino, {\it et al.}, J. Phys. Soc. Jpn. 68, 1496 (1999); cond-mat/9902048. 


\end{references}
\end{document}